\documentclass{article}
\usepackage{arxiv}

\usepackage[hidelinks]{hyperref}
\hypersetup{
  colorlinks   = true, %Colours links instead of ugly boxes
  urlcolor     = blue, %Colour for external hyperlinks
  linkcolor    = blue, %Colour of internal links
  citecolor   = teal %Colour of citations
}

\usepackage[utf8]{inputenc} %  allow utf-8 input
\usepackage[T1]{fontenc}    %  use 8-bit T1 fonts
\usepackage{url}            %  simple URL typesetting
\usepackage{booktabs}       %  professional-quality tables
\usepackage{amsfonts}       %  blackboard math symbols
\usepackage{nicefrac}       %  compact symbols for 1/2, etc.
\usepackage{microtype}      %  microtypography
\usepackage{lipsum}		    %  Can be removed after putting your text content
\usepackage{graphicx}
\usepackage{natbib}
\usepackage{doi}
\usepackage{xcolor}

\usepackage{amssymb}
\usepackage{amsmath}  
\usepackage{multirow} 
\usepackage{geometry}
\usepackage[shortlabels]{enumitem}
\usepackage{amsfonts}
\usepackage{eurosym}
\usepackage{a4}
\usepackage{epsf}
\usepackage{rotating}
\usepackage{tabularx}
\usepackage{lineno}
\usepackage[ruled,linesnumbered]{algorithm2e}
\usepackage{algorithmicx,algpseudocode}
\usepackage{subfig}
\usepackage{caption}
\usepackage{float}
\usepackage{ragged2e}

\newtheorem{definition}{Definition}

\title{Depth-based clustering analysis of directional data}

\author{ \href{https://orcid.org/0000-0000-0000-0000}{\hspace{1mm}Giuseppe Pandolfo}\thanks{ } \\
	Department of Economics and Statistics \\
	University of Naples Federico II \\
	Napoli, Italy \\
	\texttt{giuseppe.pandolfo@unina.it} \\
	\And
	\href{https://orcid.org/0000-0000-0000-0000}{\hspace{1mm}Antonio D'Ambrosio} \\
	Department of Economics and Statistics \\
	University of Naples Federico II \\
	Napoli, Italy \\
	\texttt{antdambr@unina.it} \\
}

\hypersetup{
pdftitle={A template for the arxiv style},
pdfsubject={q-bio.NC, q-bio.QM},
pdfauthor={David S.~Hippocampus, Elias D.~Striatum},
pdfkeywords={First keyword, Second keyword, More},
}

\begin{document}
\maketitle

\begin{abstract}
A new depth-based clustering procedure for directional data is proposed. Such method is fully non-parametric and has the advantages to be flexible and applicable even in high dimensions when a suitable notion of depth is adopted. The introduced technique is evaluated through an extensive simulation study. In addition, a real data example in text mining is given to explain its effectiveness in comparison with other existing directional clustering algorithms.\end{abstract}

% keywords can be removed

%----------------------------------------------------------------------------------------
%	ARTICLE CONTENTS
%----------------------------------------------------------------------------------------

\section{Introduction}
\label{intro}

Clustering is a fundamental subject in statistics which has been widely investigated in classical multivariate analysis over the past decades. It aims at organizing a set of objects into homogeneous groups, such that objects in the same cluster (or group) are more ``similar'' to each other than to objects in different clusters.  Because of its practical importance, clustering has received a lot of attention in various branches of statistics. However, less attention has been paid to the cluster analysis of directional data even though it is becoming more and more important in modern applications where observations are represented by angles or unit vectors. A few examples are the wind direction (meteorology), the orientation of magnetic fields in rocks (geology) and the  movement of animals (biology). In the recent years, thanks to new visualization techniques \citep[see][]{buttarazzi2018}, they have also seen a significant and growing interest by neuroscientists (to study the direction of neuronal axons and dendrites) and microbiologists (to analyze the angles formed by protein structures).
 
\sloppy Directional data are constrained to lie on the surface of the unit $\left(d-1\right)$-dimensional hypersphere $\mathbb{S}^{d-1}:= \left\{x \in \mathbb{R}^{d}: \left\|x\right\|_{2} = 1\right\}$, where $\left\|x\right\|_{2}:= \sqrt{\sum_{i=1}^{d}x^{2}_{i}}$, with $x = (x_{1},\ldots,x_{d})'$. Within the literature, they are presented and discussed by \cite{Mardia2000}, which is a classical reference on the subject. More recent developments of this branch of statistics can be found in \cite{ley2017modern, ley2018applied}.

Analyzing and describing directional data requires tackling some interesting problems associated with the lack of a reference direction and with a sense of rotation not uniquely defined. One more important issue is the lack of a natural ordering, which generates a special interest in depth functions on the (hyper)sphere.

Such peculiar features make the use of classical statistical methods inappropriate, and often misleading. In this regard, consider the angles $0$ and $2\pi$ on a unit circle $\mathbb{S}^{1}$. Their arithmetic mean is $\pi$, but they are actually the same angle and the ``true'' mean is $0$. Thus, working with directional data requires specific techniques that consider the geometry of the manifold, and this obviously holds true also for clustering issues. 

Despite the theory on clustering methods based on depth functions in $\mathbb{R}^{d}$ has been recently established and it is still relatively young and under development (see e.g. \citealp{Jornsten2004}, \citealp{ding2007} and \citealp{jeong2016}), to the best of authors' knowledge, there is no work of such type to perform clustering of high dimensional directional data in the literature. However, angular depth functions were recently employed to perform classification of hyper-spherical objects (see \citealp{pandolfo2021}). 

The idea of data depth, a measure of how deep or outlying a given multidimensional point is with respect to a dataset, allows generalizing the concepts of median and rank to multivariate data. This way, a multidimensional center-outward order (similar to that of a real line) can be obtained. Obviously, this holds true also for directional data. 

Depth-induced ordering enables a description of multivariate distributions and aids in using depth functions for clustering as shown in \cite{Jornsten2004} and \cite{torrente2021} who have proven power of depth function methodology. Then, even though the use of a depth notion in the directional data clustering problem is yet a new and unexplored tool, it is highly appealing to extend these ideas to such  complex setting of data. 

Hence, in this paper we propose a non-parametric method for performing cluster analysis of directional data based on the concept of data depth for directional data. The proposal aims to be considered as an alternative tool in the general framework of clustering methods for directional data.

The paper is organized as follows. In Section \ref{sec:prelim} the von Mises-Fisher distribution, the reference distribution in directional data analysis, is briefly recalled. In Section \ref{sec:depths}, we provide a brief review of angular data depths available within the literature. In Section \ref{sec:algorithm} the depth-based clustering procedure is presented. Section \ref{sec:SimulatedExamples} contains an evaluation of the proposed method through a simulated study. A real data applications for textual data analysis is presented in Section \ref{sec:RealDataExampleTextClustering}. Finally, Section \ref{sec:conclusions} contains some concluding remarks.

%--------------------------------------------------------------------------------------------------------------------------

\section{Preliminaries: The von Mises-Fisher distribution}
\label{sec:prelim}

\begin{figure}[h]
    \centering
\subfloat[$\kappa = 0$]{\includegraphics[width=0.32\columnwidth]{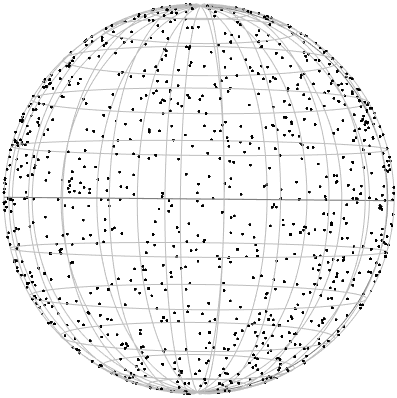}}\hfil
\hspace{-0.5cm}
\subfloat[$\kappa = 5$]{\includegraphics[width=0.32\columnwidth]{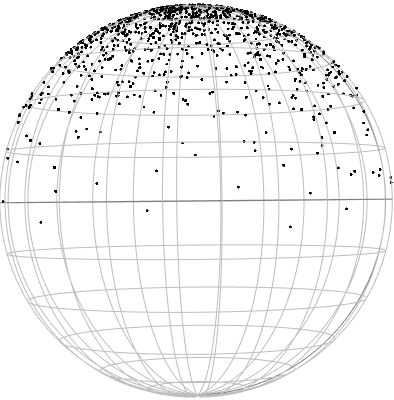}}

\subfloat[$\kappa = 20$]{\includegraphics[width=0.32\columnwidth]{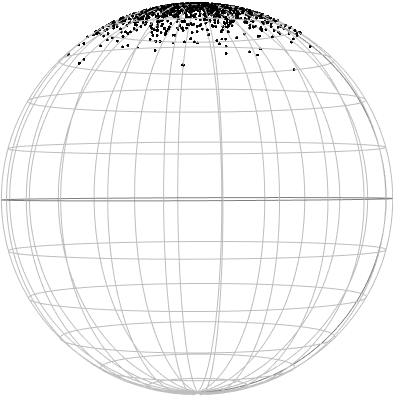}}\hfil
\hspace{-0.5cm}
\subfloat[$\kappa = 100$]{\includegraphics[width=0.32\columnwidth]{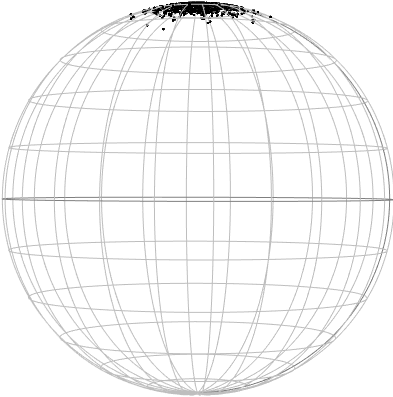}}
\caption{Plots of four samples of 1000 data points drawn from a von Mises-Fisher distribution on $\mathbb{S}^{2}$ for concentration parameter $\kappa$ equal to $0$ (a), $5$ (b), $20$ (c) and $100$ (d).}
\label{fig:vmfconc}
    \end{figure}

In this section, we briefly review the von Mises-Fisher (vMF) distribution, which arises naturally for data distributed on the unit hyper-sphere. A $\left(d-1\right)$-dimensional unit random vector $x$ (i.e., $x \in \mathbb{R}^{d}$ and $\left\|x\right\| = 1$, or equivalently $x \in \mathbb{S}^{d-1})$ is said to have a von Mises-Fisher distribution if its probability density function is given by
\begin{equation*}
f(x|\mu,\kappa) = c_{d}(\kappa)\exp\left\{\kappa \mu' x\right\},
\label{eq:vmf}
\end{equation*}
where $\left\|\mu \right\| = 1$, $\kappa \geq 0$ and $d \geq 2$. The normalizing constant $c_{d}(\kappa)$ is given by
\begin{equation*}
c_{d}(\kappa) = \frac{\kappa^{d/2-1}}{(2\pi)^{d/2}I_{d/2-1}(\kappa)},
\label{eq:}
\end{equation*}
where $I_{\nu}(\cdot)$ represents the modified Bessel function of the first kind and order $\nu$. The density $f(x|\mu, \kappa)$ is parametrized by the mean direction $\mu$, and the concentration parameter $\kappa$, so-called because it characterizes how strongly the unit vectors drawn according to $f(x|\mu, \kappa)$ are concentrated about $\mu$. Larger values of $\kappa$ imply stronger concentration about the mean direction. In particular when $\kappa = 0$, $f(x|\mu, \kappa)$ reduces to the uniform density on $\mathbb{S}^{d-1}$, and as $\kappa \rightarrow \infty$, $f(x|\mu, \kappa)$ tends to a point density. Such distribution is a reference one for directional data, and has properties analogous to those of the multivariate Normal distribution for data in $\mathbb{R}^{d}$. For example,
the maximum entropy density on $\mathbb{S}^{d-1}$ subject to the constraint that $E\left[x\right]$ is fixed is a vMF density. Figure \ref{fig:vmfconc} illustrates the impact of $\kappa$, by presenting four samples of 1000 points on the unit sphere (i.e $d = 3$) according to four vMF distributions with the same mean direction, $\mu = (0,0,1)'$, but with different values of $\kappa \in \left\{0,5,20,100\right\}$.

%--------------------------------------------------------------------------------------------------------------------------

\section{Data depth for directional data}
\label{sec:depths}

In this section, we recall the notions of data depth for directional data available within the literature, that is the angular simplicial depth, the angular halfspace depth, the arc distance depth, the cosine distance depth, and the chord distance depth. The first three were introduced and investigated by \citet{Liu1992}, while the latter by \citet{pandolfo2018distance}. 

\noindent \begin{definition}{\textbf{Angular simplicial depth (ASD).}} The angular simplicial depth of a given point $x \in \mathbb{S}^{d-1}$ w.r.t. the distribution $F$ on the unit hypersphere $\mathbb{S}^{d-1}$ is defined as follows:
\begin{equation*}
	ASD\left(x, F\right) := P_{F} \left(x \in \Delta \left(W_{1},\ldots,W_{d}\right)\right),
\end{equation*}
where $P_{F}$ denotes the probability content w.r.t. the distribution $F$. $W_{1}, \ldots, W_{d}$ are i.i.d. observations from $F$ and $\Delta \left(W_{1}, \ldots, W_{d}\right)$ denotes the simplex with vertices $W_{1}, \ldots, W_{d}$. 
\end{definition} 

\noindent \begin{definition}{\textbf{Angular Tukey's depth (ATD).}} The angular Tukey's depth of a given  point $x \in \mathbb{S}^{d-1}$ w.r.t. the distribution $F$ on $\mathbb{S}^{d-1}$ is defined as follows:
\begin{equation*}
	ATD\left(x, F\right) := \inf_{HS:x \in HS} P_{F} \left(HS \right),
\end{equation*}
where $HS$ denotes the set of all closed hemispheres containing $x$.
\end{definition}

\noindent \begin{definition}{\textbf{Arc distance depth (ADD).}} The arc distance depth of of a given point $x \in \mathbb{S}^{d-1}$ w.r.t. the distribution $F$ on $\mathbb{S}^{d-1}$ is defined as follows:
\begin{equation*}
	ADD\left(x, F\right) := \pi - \int \ell \left(x,\varphi\right) dF\left(\varphi\right),
\end{equation*}
where $\ell \left(x,\varphi\right)$ is the Riemannian distance between $x$ and $\varphi$ (i.e. the length of the shortest arc joining $x$ and $\varphi$). 
\end{definition}

\noindent \begin{definition}{\textbf{Cosine distance depth (CDD).}} The cosine distance depth of of a given point $x \in \mathbb{S}^{d-1}$ w.r.t. the distribution $F$ on $\mathbb{S}^{d-1}$ is defined as follows:
\begin{equation*}
	CDD\left(x, F\right) := 2 - E_{F}[1-x' W],
\end{equation*}
where $E_{F}$ is the expectation under the assumption that $W$ has distribution $F$. 
\end{definition}

\noindent \begin{definition}{\textbf{Chord distance depth (ChDD).}} The chord distance depth of of a given point $x \in \mathbb{S}^{d-1}$ w.r.t. the distribution $F$ on $\mathbb{S}^{d-1}$ is defined as follows :
\begin{equation*}
	ChDD\left(x, F\right) := 2 - E_{F} \left[\sqrt{2\left(1 - x'W\right)}\right],
\end{equation*}
where $E_{F}$ is the expectation under the assumption that $W$ has distribution $F$. 
\end{definition}

For convenience, the notation $AD(\cdot,\cdot)$ will be used in the following to denote any angular depth function, unless a particular notion is adopted. 

All the depth functions listed above possess the following important properties:
\begin{itemize}%[leftmargin=7.7mm]
	\item[\textbf{P1.}] \textbf{Rotation invariance}: $AD\left(x, F \right) = AD\left( Ox, OF \right)$ for any $d \times d$
orthogonal matrix $O$;
	\vspace{0.6em}
	\item[\textbf{P2.}] \textbf{Maximality at center}: $\underset{x \in \mathbb{S}^{d-1}}{\max} AD\left(x,F\right) = AD\left(x_{0},F\right)$ for any $F$ with center at $x_{0}$;
	\vspace{0.6em}
\item[\textbf{P3.}] \textbf{Monotonicity on rays from the deepest point}: $AD\left(\cdot,\cdot\right)$ decreases along any geodesic path $t \mapsto x_{t}$ from the deepest point $x_{0}$ to the antipodal point $-x_{0}$. 
\end{itemize}
In addition, one more important property is satisfied by the distance-based depths, that is:
\begin{itemize}%[leftmargin=7.7mm]
	\item[\textbf{P4.}] \textbf{Minimality at the antipodal point to the center}: $AD\left(-x_{0}, F\right) = \linebreak\underset{x \in \mathbb{S}^{d-1}}{\inf} AD\left(x, F\right)$ for any $F$ with center at $x_{0}$.  
\end{itemize}

One further available notion of depth for directional data is the angular Mahalanobis depth of \citet{ley2014}, which is based on a concept of directional quantiles. However, it requires a preliminary choice of a spherical location and for this reason it is not considered in this work. Moreover, for the purpose of this work, the ASD and ATD will not be considered. This is because they have two main drawbacks that make them not feasible for the application of the proposed method to high-dimensional data, that is: \textsl{(i)} The are high computationally demanding and \textsl{(ii)} can take zero values \citep[see][]{Liu1992}, while distance-based directional depths take positive values everywhere on $\mathbb{S}^{d-1}$ (but in the uninteresting case of a point mass distribution).

Note that depth functions are not to intended  to be equivalent with density. 
However, contours of depth are often used to reveal the shape and the structure of a multivariate data set. Such contours are analogous to quantiles in the univariate case, and they allow for the computation of $L$-estimators of location-scatter parameters (such as trimmed means). 

Unlike the univariate case, multivariate ordering can be defined in different ways, but it is usually desirable for samples from a certain class of distributions such as the rotationally symmetric ones, the angular depth contours should track the contours of the underlying model (and thus circularly contoured). Such contours thus are formed by a $(d - 1)$-dimensional sphere (a circle inside the sphere $\mathbb{S}^{2}$, two points on the circle $\mathbb{S}^{1}$) centered at $x_{0}$. Rotationally symmetric distributions are characterized by densities of the form
\begin{equation}
x \mapsto f_{x_{0}}\left(x\right) = c_{\kappa, f} f \left(x'x_{0}\right), \quad x \in \mathbb{S}^{d-1},
\label{eq:densvmf}
\end{equation}
where $f: \left[-1,1\right] \mapsto \mathbb{R}^{+}_{0}$ is an absolutely continuous and monotone increasing function and $c_{\kappa, f}$ a normalizing constant. Such class of distributions contains the von Mises-Fisher distribution which is obtained by taking $f(u) = \exp(\kappa u)$ for some strictly positive concentration parameter $\kappa$. Note that Theorems 3 and 4 in \citealp{pandolfo2018distance} ensures that the maximal depth, $\underset{x \in \mathbb{S}^{d-1}}{\max} AD_{d(\cdot)}(x, F)$, measures the concentration of $F$, irrespective of the chosen distance measure $d(\cdot)$ for distributions having density of the form given in (\ref{eq:densvmf}).  

It is usually convenient to treat depth contours in terms of their corresponding $\alpha$-regions that, as usual, are defined as the collection
of $x$ values with a depth larger than or equal to $\alpha$.
\begin{definition}
For a given angular depth function $AD\left(x,F\right)$ and for $\alpha > 0$, we call
\begin{equation*}
AD^{\alpha}\left(F\right) \equiv \left\{x \in \mathbb{S}^{d-1}|AD\left(x,F\right) \geq \alpha\right\}
\end{equation*}
the corresponding $\alpha$-depth region and its boundary $\partial AD^{\alpha}\left(x,F\right)$ the corresponding $\alpha$-contour. 
\end{definition}

The $\alpha$-depth regions for data on $\mathbb{S}^{d-1}$ usually are desired to be rotationally invariant and nested.  
In addition, it is worth underlying that the $\alpha$-regions inbuced by ADD, CDD and ChDD are invariant under rotations when fixing $x_{0}$, hence they are able to reflect the symmetry of the distribution $F$ about $x_{0}$. Roughly speaking, each depth region can be considered a sort of measure of the distance of a given point from a central point of the distribution, where the depth function takes its maximum. Hence, more dispersed data lead to larger regions.  

\begin{figure}[H]
\begin{minipage}{.5\linewidth}
\centering
\subfloat[]{\label{main:a}\includegraphics[scale=.28]{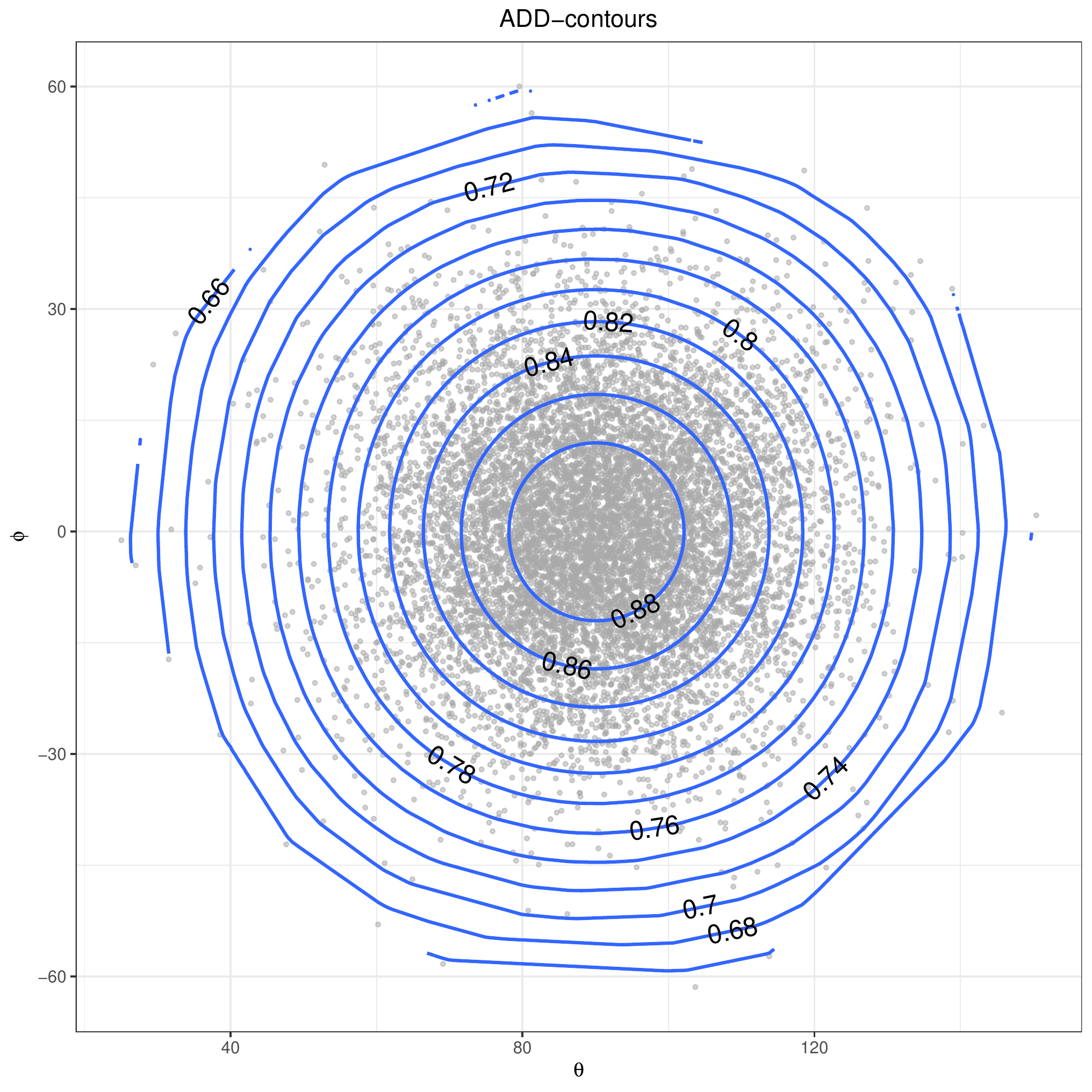}}
\end{minipage}%
\begin{minipage}{.5\linewidth}
\centering
\subfloat[]{\label{main:b}\includegraphics[scale=.28]{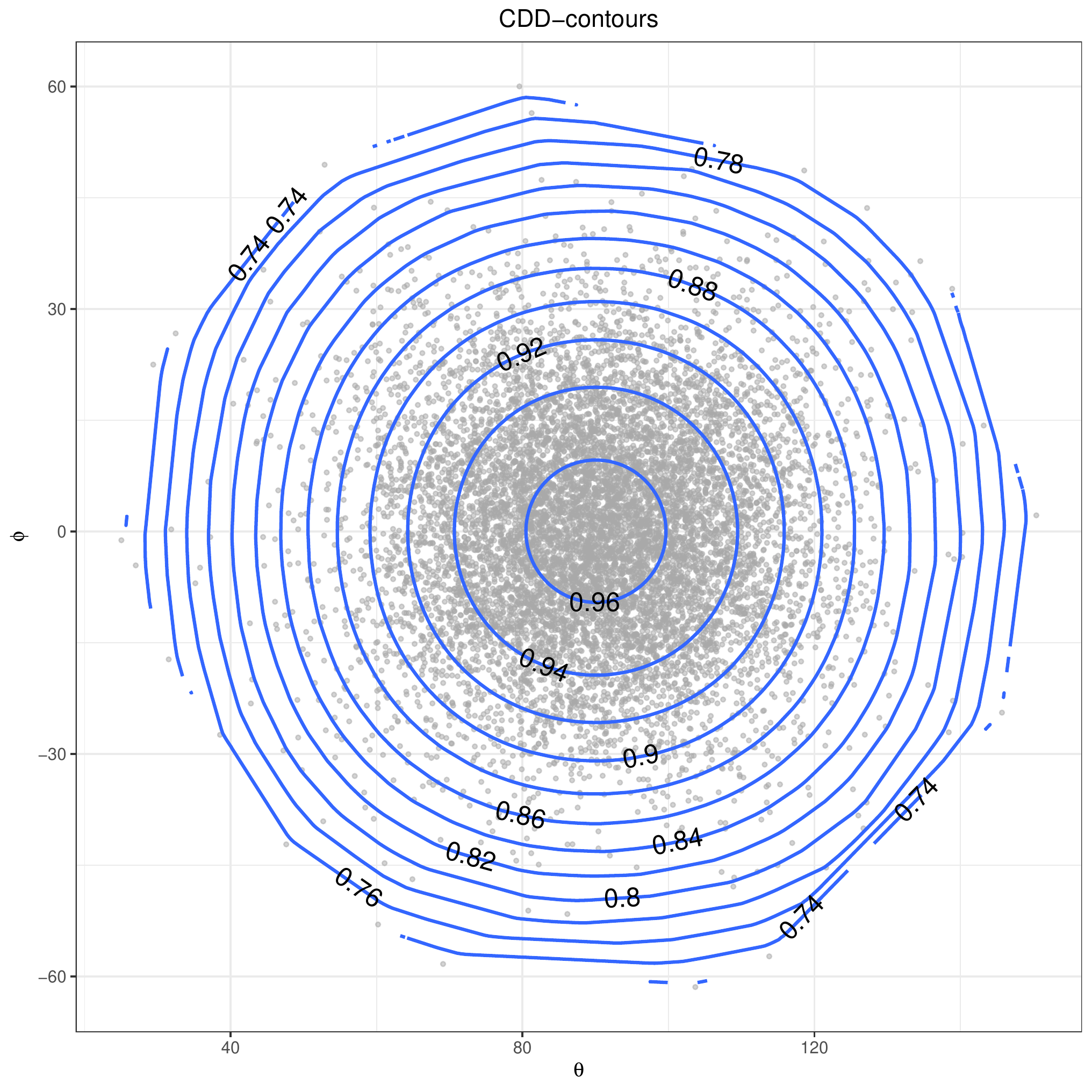}}
\end{minipage}\par\medskip
\centering
\subfloat[]{\label{main:c}\includegraphics[scale=.28]{{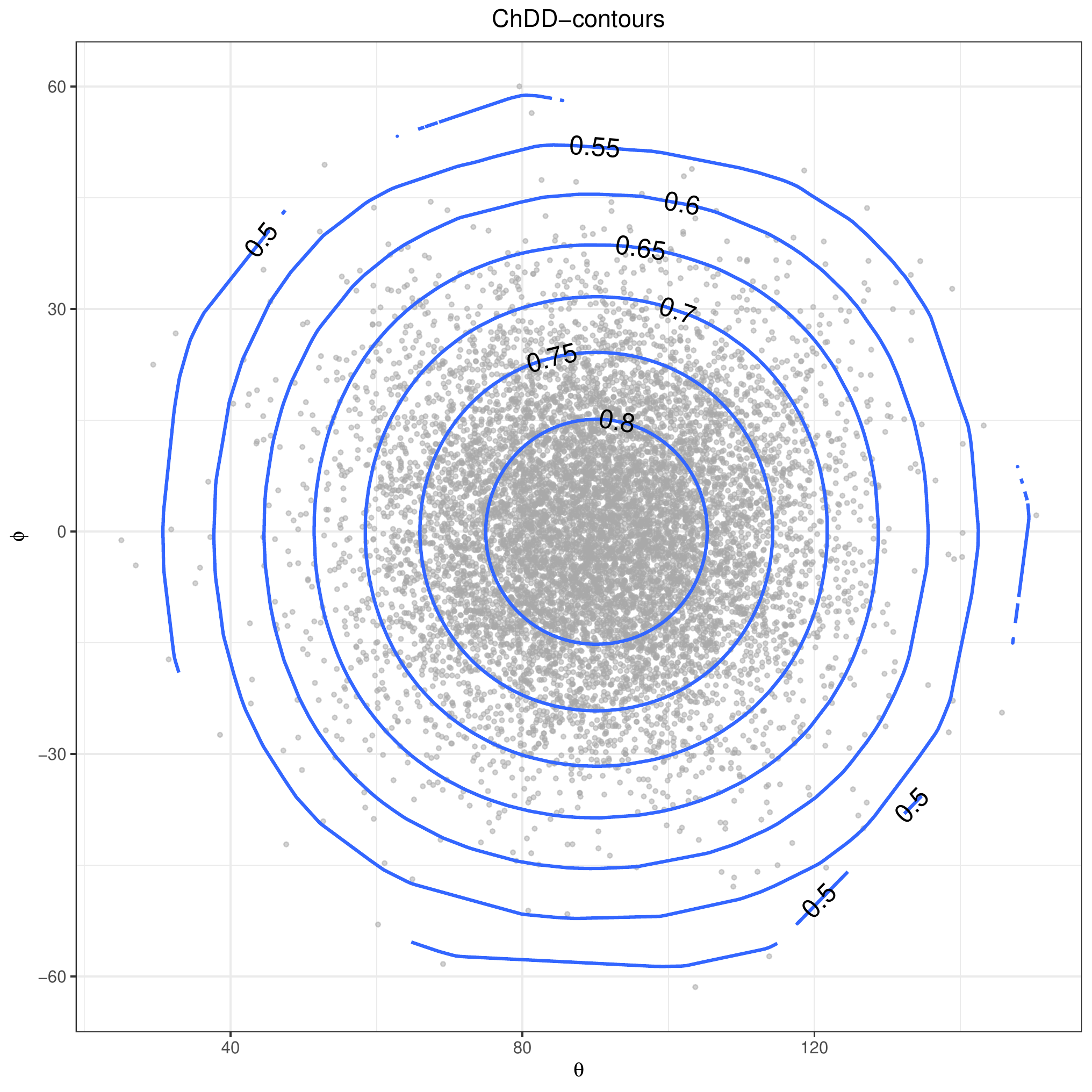}}}
\caption{Plots of the ADD (a), CDD (b) and ChDD (c) empirical contours of a sample of 1000 data points drawn from a von Mises-Fisher distribution on $\mathbb{S}^{2}$ with concentration parameter $\kappa = 15$.}
\label{fig:depthcontours}
\end{figure}

To illustrate the angular depth ordering and its ramifications, the normalized ADD, CDD and ChDD  are used to order sample points and graph their corresponding representative contours. Applying depth ordering to a sample of 1000 points drawn from a von Mises-Fisher distribution on $\mathbb{S}^{2}$ with concentration parameter $\kappa = 15$, the sample $p^{th}$ level contours in Figure \ref{fig:depthcontours} were obtained. For the sake of the illustration, data are presented in spherical coordinates using angles $\theta$ and $\phi$ with unit radius. As one can see the contours are nested within one another.

\section{Clustering directional data: a brief review}
\label{sec:overview}

The last few decades have witnessed an increasing interest in classification methods for directional data in a broad sense (which includes both classification and clustering techniques).

The literature on supervised classification of directional data is quite rich. \cite{sengupta2005simple} proposed a discrimination rule based on the chord distance. \cite{ackermann1997note} adapted discriminant analysis procedures for linear data to the analysis of circular data, while a comparison of different classification rules on the unit circle was performed by \cite{tsagris2019comparison}. The Naive Bayes classifier for directional data was introduced by \cite{lopez2015directional}, and the discriminative directional logistic regression by \cite{fernandes2016discriminative}. More recently, \cite{di2019kernel} considered non-parametric circular classification based on Kernel density estimation. \cite{pandolfo2021} and \cite{Demni2021} studied the depth-versus-depth (DD) classifier for directional data, while \cite{demni2019cosine} introduced a cosine depth based distribution method. 

The development of clustering techniques for directional data has recently been an important research topic. The two most popular approaches to perform clustering of data on $\mathbb{S}^{d-1}$ are the spherical $k$-means \citep{dhillon2001concept} and the use of mixture models with vMF components. The former aims at maximizing the cosine similarity $\sum_{i=1}^{n}X_{i}^{'}c_{i}$ between the sample $X_{1},\ldots,X_{n}$ and $k$ centroids $c_{1},\ldots,c_{k} \in \mathbb{S}^{d-1}$, where $c_{i}$ is the centroid of the cluster containing $X_{i}$. \cite{dhillon2003modeling} and \cite{banerjee2003generative,banerjee2005} gave Expectation-Maximization (EM) algorithm for fitting vMF mixtures which have spherical $k$-means as a particular case. Other different approaches for fitting vMF mixtures were proposed by \cite{yang1997fuzzy}, based on embedding fuzzy c-partitions in the mixtures, and \cite{taghia2014bayesian} who addressed the Bayesian estimation of the vMF mixture model employing variational inference. \cite{franke2016mixture} developed an EM algorithm to fit general projected normal mixtures on $\mathbb{S}^{2}$.

A fuzzy variation of the $k$-means clustering procedure for directional data was proposed by \cite{kesemen2016}, while \cite{possibilistic2019} introduced possibilistic c-means.

\section{Depth-based medoids clustering algorithm}
\label{sec:algorithm}

In this section a depth-based medoids clustering algorithm (DBMCA) for directional data is introduced. Some concepts which are required for the formulation of the proposed algorithm are defined below. 

\begin{definition}\textbf{(Depth-based partition)}: A depth-based partition $\mathcal{C}_k$ is a non-empty maximal and depth-based subset of a $n \times (d-1)$ data set $\mathbf{X}$ defined in $\mathbb{S}^{d-1}$ such that $\mathbf{X} = \left\{\mathcal{C}_1,\ldots,\mathcal{C}_k,\ldots \mathcal{C}_{K}\right\}$, where $K \leq n$. 
\end{definition}

\begin{definition} \textbf{(Depth-medoid)}: A depth-medoid $x_{\scriptscriptstyle AD_{k}} \in \mathbb{S}^{d-1}$ is a data point belonging to the depth-based partition $\mathcal{C}_k$ for which
\[
x_{\scriptscriptstyle AD_{k}} = \underset{\mathbf{X} \in \mathcal{C}_{k}}{\text{argmax}}~AD\left(\mathbf{X} \right).
\]
Hence, the depth-medoid is the point with the largest depth value within the $k^{th}$ depth-based partition.%, representing its \emph{prototype}.
\end{definition}

One further necessary ingredient in cluster analysis is a similarity measure to quantify how ``close'' objects are to each other. To this end a depth function appears to be useful since it quantifies how closely an observed point is located to the ``center'' of a finite set, or relative to a probability distribution. Hence, to this end we propose to use any notion of angular statistical data depth as an alternative similarity measure. That is, let the data be stored as an $n \times (d - 1)$ matrix $\mathbf{X}$. Then the depth-based similarity can be defined as:
$$
sim_{\scriptscriptstyle AD}(x_{i},x_{j}) = AD(x_{i},x_{j}), \quad \text{with} \quad i \neq j.
$$
Then we use such a depth-based similarity measure for searching the closest neighbors.

\begin{definition} \textbf{(Depth-medoid neighbors)}: A depth-medoid neighbor is a data point $x_{i}$ belonging to the depth-based partition $\mathcal{C}_{k}$, with medoid $x_{\scriptscriptstyle AD_{k}}$, for which
\[
sim_{\scriptscriptstyle AD}\left(x_{i}, x_{\scriptscriptstyle AD_{\scriptscriptstyle k}}\right)  > sim_{\scriptscriptstyle AD}\left(x_{i}, x_{\scriptscriptstyle AD_{\scriptscriptstyle j}}\right) \quad \text{with}~k \neq j.
\]
In other words a depth-medoid neighbor of the partition $\mathcal{C}_{k}$  is a point whose depth-based similarity w.r.t. the depth-medoid of the $k^{th}$ partition is maximum of all the clusters $\mathcal{C}$. This is 
\end{definition}

Hence, in order to evaluate the goodness of the partition into $k$ clusters, the proposed algorithm makes use of a depth-based cluster homogeneity measure. As highlighted in \cite{hoberg2000}, data showing larger dispersion give rise to larger $\alpha$-depth regions, and consequently, homogeneity can be measured by considering the depth values of data points within regions. More in detail, the idea is to use a \textit{within-class depth-based concentration measure} by following the depth-based dispersion measure proposed by \cite{romanazzi2009data} for data in $\mathbb{R}^{d}$ and later used by \cite{agostinelli2013nonparametric} for the analysis of directional data. Specifically, the depth-based homogeneity within the $k^{th}$ cluster can be defined as the expected measure of the angular depth within each cluster $\mathcal{C}_{k}$
\begin{equation*}
DW_{k} := \int_{\mathcal{C}_{k}}AD_{\scriptscriptstyle \mathbf{X}}\left(x\right)d \upsilon \left(x\right),
\end{equation*}
where $\upsilon$ denotes the Lebesgue measure on $\mathcal{C}_{k}$.
Such measure can be approximated for sample data as
\begin{equation}
	%DW_{k}\left(\left\{\mathcal{C}_{j},\ldots, \mathcal{C}_k, \ldots, \mathcal{C}_{K}\right\}\right) := \sum_{j \in \mathcal{C}_{k}}AD^{k}_{i},
	\widetilde{DW}_{k}:=  \frac{1}{n_{\scriptscriptstyle k}}\sum^{n_{\scriptscriptstyle k}}_{i = 1}AD_{n}(x_{i}),
\label{eq:dw}
\end{equation}
where $n_{k}$ denotes the size of the $k^{th}$ cluster. 

The algorithm searches for the partition of the data points into $k$ clusters in such a way that the expected angular depth of each cluster is maximized.

The proposed algorithm follows the well-known \emph{partitioning around medoids} (PAM) approach \citep{pam}. The main difference is that we iteratively search for those points, the depth-medoids, that maximize a given depth function between themselves and other points belonging to the same partition.\\
More formally, let $\mathbf{X}$ be a $n \times (d-1)$ data set defined in $\mathbb{S}^{d-1}$, randomly select $K$ observations as depth-medoids. Iteratively:
\begin{itemize}
\item Assign each data point to the cluster for which the angular depth function $AD\left( \cdot \right)$ between itself and the $K$ depth-medoids is maximum (assignment step);
\vspace{0.75em}
\item Refine the $K$ depth-medoids by choosing for each cluster those points that allow for the maximization of the depth-based within-cluster homogeneity as defined in (\ref{eq:dw}) (refinement step).
\end{itemize}

Here, for evaluating the clustering results and determining the optimal number of clusters, the Silhouette index of \cite{rousseeuw87} is adopted. It defines for each object in a data set, the measure of how this object is similar to other objects from the same cluster (cohesion) in comparison with objects of other clusters (separation).  
Specifically, in our approach, data consists of angular depth-based \textit{similarities}. 
Hence, assume the data have been clustered into $k$ clusters, then for each vector $i \in \mathcal{C}_{i}$, we have 
\begin{align*}
a'(i) &= \text{average~depth-based~similarity~of~$i$~to~all~other~points~in~its~own~cluster~$\mathcal{C}_{i}$} \\
d'(i) &= \text{average~depth-based~similarity~between~$i$~to~the~objects~in~any~other~cluster} \\
b'(i) &= \underset{j \neq i}{\max}~d'(i, \mathcal{C}_{j}) 
\end{align*}
then, with such measure, the silhouette coefficient is defined as:
\begin{equation}
  s(i) =
    \begin{cases}
      1-b'(i)/a'(i) & \text{if}~a'(i) > b'(i) \\
      0                  & \text{if}~a'(i) = b'(i) \\
      a'(i)/b'(i)-1 & \text{if}~a'(i) < b'(i)
    \end{cases}       
\end{equation}

%--------------------------------------------------------------------------------------------------------------------------

\subsection{Initialization of cluster centers}
\label{sec:initclustcenters}

Initialization of iterative algorithms can have a significant impact on the resulting performances.  Several techniques can be found within the literature. Here, we propose a modified version of the well-known \textit{\texttt{k-means++}} algorithm proposed by \cite{init}.  
Specifically, assuming the number of clusters is equal to $k$ and denoting the depth-based similarities by $sim_{\scriptscriptstyle AD}$: 
\begin{description}%[leftmargin=2.7\parindent]
\setlength\itemsep{0.7em}
\item[\textbf{Step 1.}] Select one data point uniformly at random from the data set $\mathbf{X}$. The chosen data point is defined as the first medoid $m_{1}$.
\item[\textbf{Step 2.}] Compute the angular data depth-based similarities between each data point (not chosen yet) and $m_{1}$ (which has already been defined). 
Denote the angular depth-based similarity of the data point $i$ w.r.t. $m_{j}$ as $sim_{\scriptscriptstyle AD}(x_{i}, m_{j})$.
\item[\textbf{Step 3.}] Choose one new data point at random as new medoid from $\mathbf{X}$ using the probability of any data point to be chosen as
$$
\frac{sim_{\scriptscriptstyle AD}(x_{i},m_{1})}{\sum_{j=1}^{n}sim_{\scriptscriptstyle AD}(x_{j},m_{1})}.
$$
\item[\textbf{Step 4.}] Then, to choose the $j^{th}$ medoid:
\vspace{0.5em}
\begin{itemize}
\setlength\itemsep{0.45em}
	\item[a.] Compute the angular depth of each data point w.r.t. each medoid, and assign each data point the medoid w.r.t. it is has maximal depth value.
	\item[b.] For $i = 1,\ldots, n$ and $p = 1, \ldots, j-1$, select the $j^{th}$ medoid at random from $\mathbf{X}$ with probability
	\begin{equation*}
	\frac{sim_{\scriptscriptstyle AD}(x_{n}, M_{p})}{\underset{h: x_{h}\in M_{p}}{\sum}(x_{h}, m_{p})},
\end{equation*}
where $M_{p}$ is the set of all observations closest to medoid $m_{p}$ and $x_{n}$ belongs to $M_{p}$.

That is, select each subsequent medoid with a probability proportional to the angular depth value of itself to the closest medoid that was already chosen.
\end{itemize}
\item[\textbf{Step 5.}] Repeat Step 4 until $k$ medoids are chosen.
\end{description}

\section{Simulation study}
\label{sec:SimulatedExamples}

We have run a a comprehensive simulation study in order to evaluate the performance of the proposed methodology. We have considered four factors, i.e. the number of clusters, the dimensionality of the problem, the level of noise, and structured vs unstructured data. We generated data with two, three, four and five theoretical partitions. We have sampled data from the von Mises-Fisher distribution in dimensions $d \in \{3, 5 ,10\}$. The level of noise was set by randomly choosing the value of the concentration parameters $k_{Low} \sim U[10,12]$, $k_{Medium} \sim U[6,8]$ and $k_{High} \sim U[2,4]$ for the cases of low, medium and high noise level, respectively.

In the case of structured data, we first selected randomly a point as center of the first partition. Then,
\begin{description}[font=$\bullet$\scshape\bfseries]
\setlength\itemsep{0.65em}
\item for the case of two centers, the second cluster center was randomly chosen with the constraint that the cosine distance from the first was between $1.7$ and $2$;
\item in the case of three centers, we randomly added a point with the constraint that the cosine distance from both the first two data points was between $0.8$ and $1.2$;
\item in the case of four centers, we randomly added a data point with the constraint that the cosine distance from the third center was between $1.7$ and $2$
\item in the case of five centers, we randomly added a data point with the constraint that the cosine distance from all other centers was between $0.5$ and $0.8$.
\end{description}
In the case of unstructured data, the theoretical centers were sampled without any constraint from a von Mises-Fisher distribution with concentration parameter $k=0$.\\
For any combination, we generated a sample of size 500. The proportion of cases was randomly chosen in such a way that the clusters were not extremely unbalanced. For each condition and for each level, ten data sets were generated, for a total of 720. Table \ref{tab:factors} summarizes the factors and the levels of the factorial design.
\begin{table}[h!]
\renewcommand{\arraystretch}{1.5}
\centering
\caption{Summary of independent factors and levels of the simulation study.}
\label{tab:factors}
\begin{tabular}{c|c}
\hline
\textbf{Factor}             & \textbf{Level}                    \\ \hline
Number of clusters & 2, 3, 4, 5               \\
Dimensions         & 3, 5, 10                 \\
Noise level        & Low, Moderate, High      \\
Data structure     & Structured, Unstructured \\ \hline
\end{tabular}
\end{table}

We adopted the cosine distance depth (CDD), the arc distance depth (ADD) and the chord distance depth (ChDD). For each trial, the number of clusters ranged from 1 to 10.  

As for each data set we know the theoretical partition, to determine how well the resulting partition matched the gold standard partition, the adjusted Rand index (ARI) of \cite{ari} which is a modified version of the Rand index which adjusts for agreement by chance. It is defined as follows.

Given an $n \times p$ data matrix $X$, where $n$ is the number of objects and $p$ the number of variables, the crisp clustering structure can be presented as a set of nonempty $K \geq 2$ subsets $\left\{C_{1},\ldots,C_{k},\ldots,C_{K}\right\}$ such that:
\begin{equation*}
X =\bigcup_{k=1}^{K}C_{k}, ~ C_{k} \bigcap C_{k'} = \emptyset, \quad \text{for}~k \neq k'
%\label{eq:}
\end{equation*}
Two objects of $X$, i.e., $(x, x')$, are paired in $C$ if they belong to the same cluster. Let $P$ and $Q$ be two partitions of $X$. The Rand index is calculated as follows:
\begin{equation*}
RI = \frac{a+d}{a+b+c+d}=\frac{a+d}{\binom{n}{2}}
\label{eq:RI}
\end{equation*}
where
\begin{itemize}
\justifying
	\item[-] $a$ is the number of pairs $(x, x') \in X$ that are paired in $P$ and in $Q$;
	\vspace{0.3em}
	\item[-] $b$ is the number of pairs $(x, x') \in X$ that are paired in $P$ but not paired in $Q$;
  \vspace{0.3em}
	\item[-] $c$ is the number of pairs $(x, x') \in X$ that are not paired in $P$ but are paired in $Q$ and
  \vspace{0.3em}
	\item[-] $d$ is the number of pairs $(x, x') \in X$ that are not paired in either $P$ or in $Q$.
\end{itemize}
The Rand index takes values in $[0, 1]$, with 0 indicating that the two partitions do not agree for any pair of elements and 1 indicating that the two partitions are exactly the same. Moreover, the Rand index is reflexive, namely, $RI(P, P) = 1$. However, such index has some drawbacks: \textit{(i)} it concentrates in a small interval close to 1, thus presenting low variability; \textit{(ii)} it approaches its upper limit as the number of clusters increases; and \textit{(iii)} it is extremely
sensitive to the number of groups considered in each partition and their density. To overcome these problems, \cite{ari} proposed the Adjusted Rand index (ARI), which can be defined as follows:
\begin{equation*}
ARI = \frac{2\left(ad-bc\right)}{b^{2}+c^{2}+2ad+(a+d)+(c+b)}.%=\frac{RI - E(RI)}{1 - E(RI)}.
\label{eq:ARI}
\end{equation*}
The ARI can yield a value between $-1$ and $+1$. ARI equal to 1 means complete agreement between two clustering results, whereas ARI equal to $-1$ means no agreement between two clustering results.

\subsection{Simulation results}
\label{simres}

\begin{figure}[h!]
	\centering
		\includegraphics[width=1.00\textwidth]{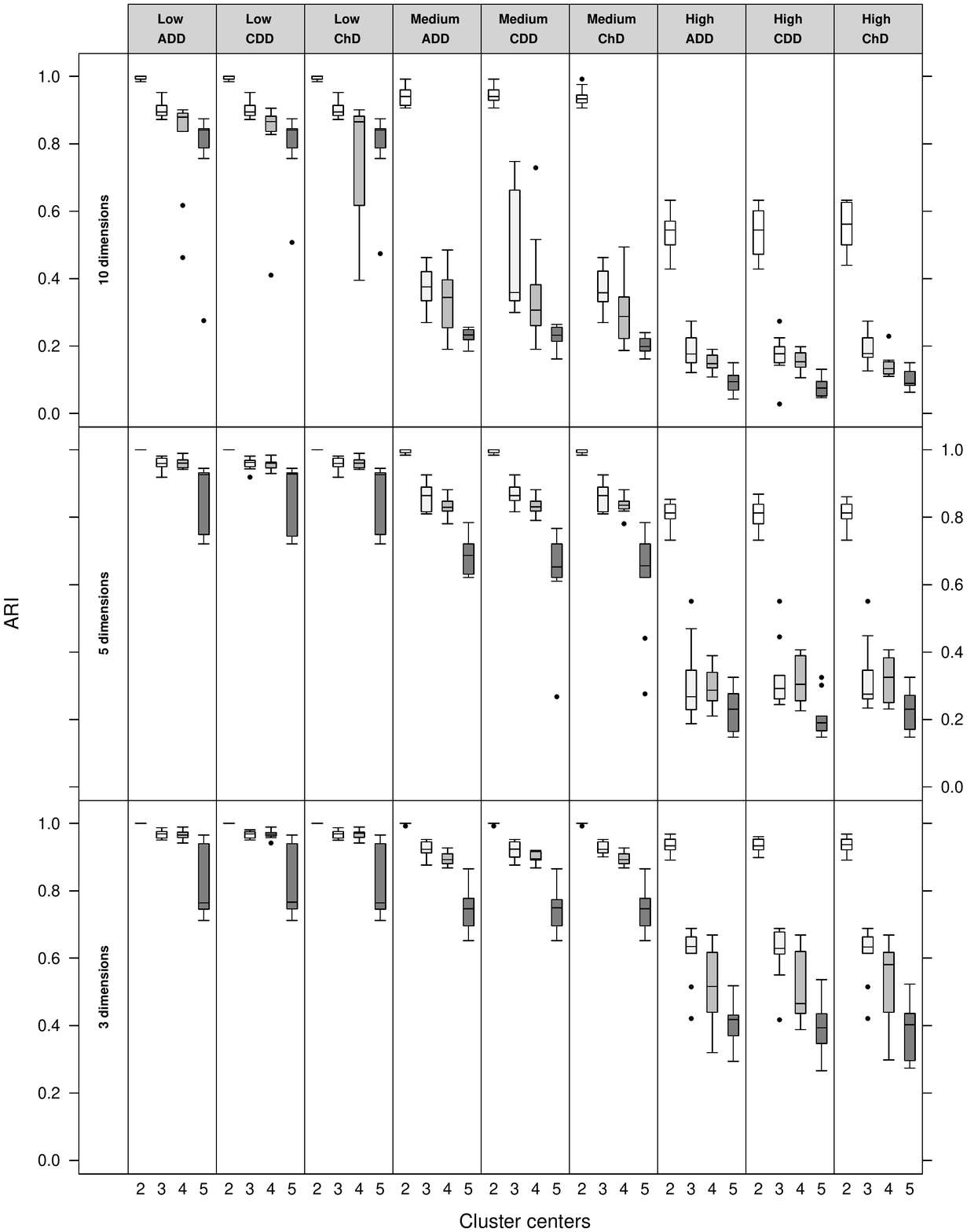}
	\caption{Structured data. The boxplots report the Adjusted Rand Index (ARI) values between the true partition and the
resulting partition given by DBMCA. %Interaction box-plots for the structured data sets. Boxes are referred to the ARI.
	}
	\label{fig:Structured}
\end{figure}

\begin{figure}[h!]
	\centering
		\includegraphics[width=1.00\textwidth]{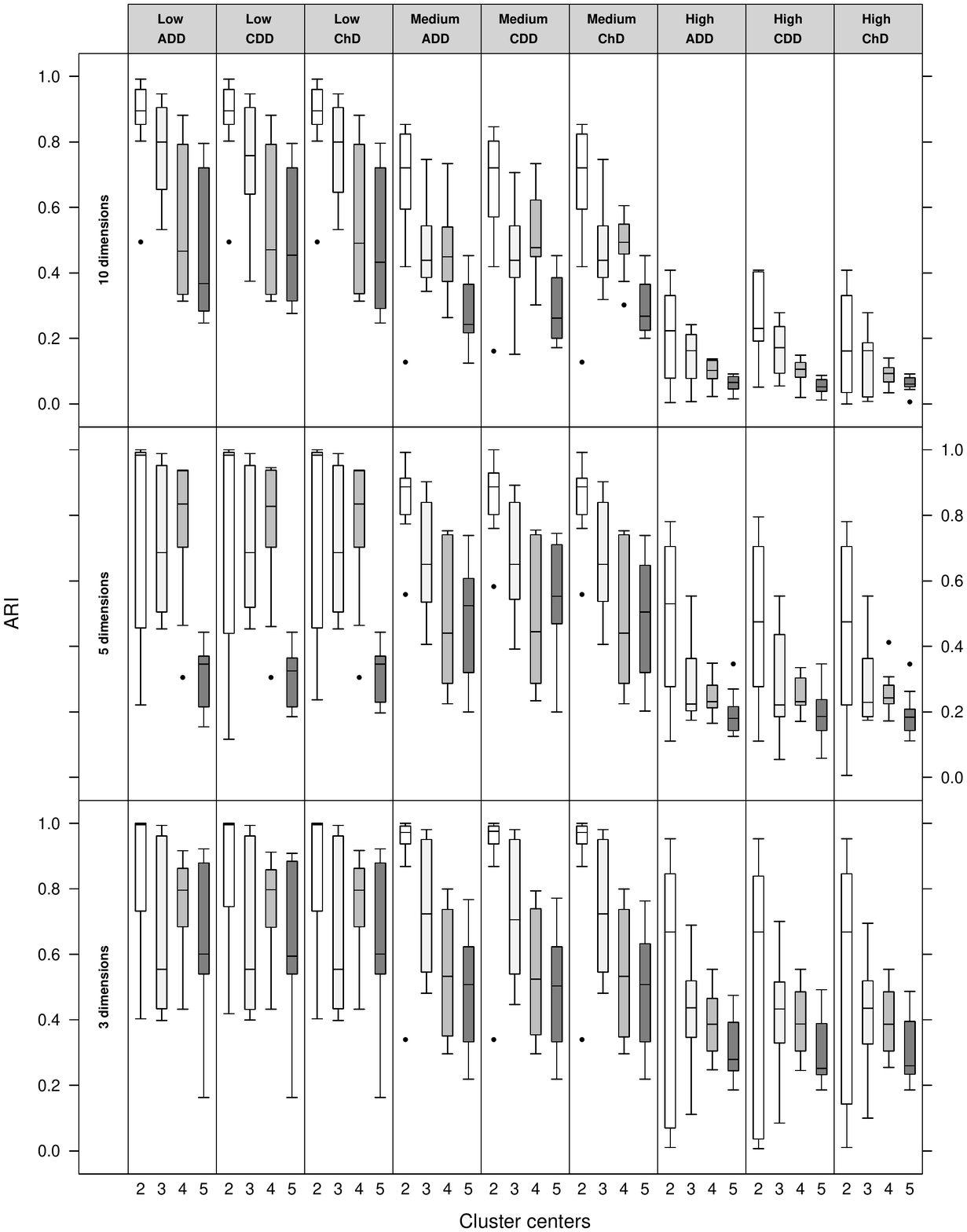}
	\caption{Unstructured data. The boxplots report the Adjusted Rand Index (ARI) values between the true partition and the
resulting partition given by DBMCA. 
}
	\label{fig:Unstructured}
\end{figure}

The results of the simulation study are presented in Figures \ref{fig:Structured} and \ref{fig:Unstructured} for structured and unstructured data, respectively. Boxplots of the ARI values of the proposed method for any combination of a dimension, a depth function, a noise level and a  number of clusters are plotted there. 

As one can see, since the performance of the depth functions under evaluation appear quite similar for both structured and unstructured data, no clear-cut indications arise on which of them is to be preferred in such cases. 

In the case structured data and for a low noise level, the ARI values show larger variability for $k=5$ clusters regardless of the dimension. When $k=2$ centers are considered, the overall variability of the ARI values is really small, and in general they are approximately equal to 0.8, except for the case of high noise level in ten dimensions where they range from 0.4 to 0.6. As expected, when a high noise is introduced to obscure the underlying clustering structure to be recovered, the ARI values get generally smaller.

In the case of unstructured data, the ARI values are generally smaller and show much more variability with respect to the case of structured data. In addition, the ARI values get smaller when the number of the theoretical clusters increases. Here again, the ARI values get generally smaller when high noise occurs. 

Internal validity values are always large for both structured and unstructured data sets, and thus not reported here.

%-----------------------------------------------

\section{Real data example: text clustering}
\label{sec:RealDataExampleTextClustering}

Directional data can arise in textual data analysis, and text clustering in particular. It has been experimentally shown that it is often helpful to normalize the data vectors to remove the bias arising from the length of a document \citep{dhillon2001}.

Here we show an example on the ``res0'' data set coming from the internal repository of the CLUTO software for clustering high-dimensional data sets (\url{http://glaros.dtc.umn.edu/gkhome/views/cluto}). It contains $1504$ documents for which the frequency of a set of $2286$ possible words have been recorded. 

The idea is modeling the word frequencies as spherical data by normalizing them to $1$ according to the $L^{2}$ norm. Normalization makes long or short documents comparable and projecting documents onto the unit hypersphere is especially useful for sparse data, as is typically the case with textual data. Indeed, this example shows a data matrix with more than the $89\%$ of zero entries.
 
For such data set, there exists also an a-priori classification of the documents, consisting of $13$ classes of documents as summarized in table \ref{tab:re0}.

\begin{table}[h!]
\renewcommand{\arraystretch}{1.5}
\centering
\caption{Frequency distribution of the a-priori partitions of the Re0 data set.}
\label{tab:re0}
\begin{tabular}{c|ccccccc}
\hline
\textbf{Class}      & housing                    & money                      & trade                      & reserves                   & cpi                        & interest                   & gnp                     \\ \hline
\textbf{Proportion} & \multicolumn{1}{r}{0.0106} & \multicolumn{1}{r}{0.4043} & \multicolumn{1}{r}{0.2121} & \multicolumn{1}{r}{0.0279} & \multicolumn{1}{r}{0.0399} & \multicolumn{1}{r}{0.1456} & \multicolumn{1}{r}{0.0530} \\ \hline \hline
\textbf{Class}      & retail                     & ipi                        & jobs                       & lei                        & bop                        & wpi                        &                            \\ \hline 
\textbf{Proportion} & \multicolumn{1}{r}{0.0133} & \multicolumn{1}{r}{0.0246} & \multicolumn{1}{r}{0.0259} & \multicolumn{1}{r}{0.0073} & \multicolumn{1}{r}{0.0253} & \multicolumn{1}{r}{0.0100} & \multicolumn{1}{r}{}      \\
\hline
\end{tabular}
\end{table}

We applied the proposed procedure to the data set by adopting the cosine distance depth (CDD). 
 We compared our proposal with the spherical $k$-means algorithm (SKM), and the mixture of von Mises-Fisher, by using both hard (movMFh) and soft (movMFs) partitioning. 
To determine the ``optimal'' number of clusters (ranging from 1 to 15), the silhouette approach was adopted for DBMCA and SKM algorithms, while the BIC criterion was used in the case of mixture of von Mises-Fisher distributions.  
A number of $k=10$ partitions was selected for DBMCA, while $k=9$ clusters were identified for spherical $k$-means. For the ``hard'' mixture of von Mises-Fisher distributions (movMFh) and its ``soft'' version (movMFs), $k=15$ and $k=14$ clusters were returned, respectively.
The spherical $k$-means and the mixture of von Mises-Fisher distributions algorithms were run through the \texttt{R} packages \texttt{skmeans} \citep{skmeans} and \texttt{movMF} \citep{movMF}, respectively. The depth-based medoids clustering algorithm (DBMCA) was computed by means of \texttt{R} functions written by the authors. 

Since the ``soft'' mixture of von Mises-Fisher distributions assigns membership probabilities of each point to each of the $k$ components (clusters), a fuzzy version of the Rand and adjusted Rand index was used, that is the normalized degree of concordance (NDC) which is defined as follows. 

Let $\mathbf{W}$ be a probabilistic (fuzzy) partition of the data matrix $\mathbf{X}$, and let $\mathbf{w}(\mathbf{x}_i) = (\mathbf{w}_1(\mathbf{x}_i)), \mathbf{w}_2(\mathbf{x}_i),\ldots, \mathbf{w}_K(\mathbf{x}_i) \in [0,1]^{K}$ be the membership degree of $\mathbf{x}_i$ in the $k$-th cluster. Given any pair $(\mathbf{x}_i, \mathbf{x}_j) \in \mathbf{X}$, \citet{huller} defined a fuzzy equivalence relation on $\mathbf{X}$ in terms of similarity measure as:
\begin{equation*}\label{equat_7}
E_{\mathbf{W}} = 1 -  \|\mathbf{W}(\mathbf{x}_i) - \mathbf{W}(\mathbf{x}_j)\|,
\end{equation*}
where $\|\cdot\|$ is the normalized $L_1$-norm, yielding a value in $[0,1]$. \\
Given two fuzzy partitions, say $\mathbf{G}$ and $\mathbf{H}$, the degree of concordance between a pair of observations is defined as
\begin{equation*}\label{equat_8}
conc(\mathbf{x}_i, \mathbf{x}_j) = 1 - \|E_{\mathbf{g}}(\mathbf{x}_i, \mathbf{x}_j) - E_{\mathbf{h}}(\mathbf{x}_i, \mathbf{x}_j)\| \vspace{3mm} \in [0,1],
\end{equation*}
yielding to a distance measure defined by the normalized sum of \textit{concordant pairs}:
\begin{equation}\label{equat_10}
d(\mathbf{G},\mathbf{H}) = \frac{1}{n(n-1)/2}\sum_{i \neq j}^n \|E_{\mathbf{g}}(\mathbf{x}_i, \mathbf{x}_j) - E_{\mathbf{h}}(\mathbf{x}_i, \mathbf{x}_j)\|,
\end{equation}
where $n$ is the sample size. The normalized degree of concordance (NDC) between two fuzzy partitions, or between a fuzzy and a crisp partition, has been defined as \citep{huller}:
\begin{equation}\label{equat_11}
NDC(\mathbf{G},\mathbf{H}) = 1 - d(\mathbf{G},\mathbf{H}).
\end{equation}
Note that it reduces to the original Rand Index when partitions $\mathbf{P}$ and $\mathbf{Q}$ are non-fuzzy.\\
The adjusted version of the NDC ``for the chance'', called adjusted concordance index, has been defined by \cite{ACI} as
\begin{equation}\label{aci}
ACI(\mathbf{G},\mathbf{H}) = \frac{NDC(\mathbf{G},\mathbf{H}) - \overline{NDC}(\mathbf{G},\mathbf{H})}{1 - \overline{NDC}(\mathbf{G},\mathbf{H})},
\end{equation}
\noindent where $\overline{NDC}(\mathbf{G},\mathbf{H})$ is the mean value of the NDC($\mathbf{G},\mathbf{H}$), which is computed by repeatedly permute one of the two partitions, say $\mathbf{\tilde{H}}$, by keeping fixed the partition $\mathbf{G}$, and then compute NDC as in Eq. (\ref{equat_11}). For an extensive overview of the fuzzy extension of the adjusted concordance index, we refer to \cite{ACI}. Both NDC and ACI are implemented in the \texttt{R} package \texttt{ConsRankClass} \citep{consrankclass}, that has been used to produce the results in the following of the paper.\\

The results summarized in Table \ref{bestbest} give the external validation indexes for all of the clustering techniques that were applied to the data set. One can note that the most similar partitions were returned by the mixtures of von Mises-Fisher distributions (ACI = 0.5391). The partitions returned by DBMCA and SKM is quite similar as well (ARI = 0.5123). The adjusted concordance index between the partitions of DBMCA and movMFs is equal to $0.3775$, which indicates a low similarity.

\begin{table}[H]
\renewcommand{\arraystretch}{1.5}
\centering
\caption{External validation measures between the partitions given by the four considered clustering techniques applied to the ``res0" data set.}
\label{bestbest}
\begin{tabular}{lc|c|c||lc|c|c}
\hline
 & \multirow{2}{*}{\begin{tabular}[c]{@{}c@{}}DBMCA\\ SKM\end{tabular}} & \multirow{2}{*}{\begin{tabular}[c]{@{}c@{}}DBMCA\\ movMFh\end{tabular}} & \multirow{2}{*}{\begin{tabular}[c]{@{}c@{}}SKM\\ movMFh\end{tabular}} & \multicolumn{1}{c}{} & \multirow{2}{*}{\begin{tabular}[c]{@{}c@{}}DBMCA\\ movMFs\end{tabular}} & \multirow{2}{*}{\begin{tabular}[c]{@{}c@{}}SKM\\ movMFs\end{tabular}} & \multirow{2}{*}{\begin{tabular}[c]{@{}c@{}}movMFh\\ movMFs\end{tabular}} \\
 &  &  &  &  &  &  &  \\ \hline
\multicolumn{1}{c}{ARI} & \multicolumn{1}{r|}{0.5123} & \multicolumn{1}{r|}{0.3918} & \multicolumn{1}{r||}{0.4583} & ACI & \multicolumn{1}{r|}{0.3775} & \multicolumn{1}{r|}{0.4824} & \multicolumn{1}{r}{0.5391} \\
\multicolumn{1}{c}{RI} & \multicolumn{1}{r|}{0.8981} & \multicolumn{1}{r|}{0.8970} & \multicolumn{1}{r||}{0.8974} & NDC & \multicolumn{1}{r|}{0.8783} & \multicolumn{1}{r|}{0.8894} & \multicolumn{1}{r}{0.9195} \\ \hline
\end{tabular}
\end{table}

Table \ref{besttrue} contains the Rand and adjusted rand indexes yielded by each method. Globally the employed clustering algorithms show quite similar results. A careful observation allows us to note that the highest ARI value is associated to DBMCA, while movFMh provides the largest RI value.

\begin{table}[H]
\renewcommand{\arraystretch}{1.5}
\centering
\caption{External validation measures with the ``true'' partition for four considered clustering techniques applied to the ``res0" data set. When hard and soft partitions are compared, the Rand index (RI) and the adjusted Rand index (ARI) are replaced with the normalized concordance index (NDC) and the adjusted concordance index (ACI), respectively.}
\label{besttrue}
\begin{tabular}{cr|r|r||cr}
\hline
\multicolumn{1}{l}{} & \multicolumn{1}{c|}{DBMCA} & \multicolumn{1}{c|}{SKM} & \multicolumn{1}{c||}{movMFh} &  & \multicolumn{1}{c}{movFMs} \\
ARI & 0.2128 & 0.2034 & 0.1854 & ACI & 0.2094 \\
RI & 0.7642 & 0.7568 & 0.7730 & NDC & 0.7659 \\ \hline
\end{tabular}
\end{table}

In addition, we computed both the ARI, ACI, RI and NDC indexes between the partitions returned by each pair of algorithms for $k \in \{1,\ldots,15\}$ whose values are reported in Table \ref{tab:ARIS}. The Rand index and, consequently, the normalized degree of concordance show always higher values than ARI and ACI. Such results were expectable since RI is not able to take into consideration effects of random groupings.

According to the ARI values, we can notice a moderate ``closeness'' of the partitions returned by the DBMCA and SKM for $k=3$, $4$, $10$, $11$ and $12$. The largest value of ACI is between DBMCA and movMFs when $k=5$. On average, the partitions returned by SKM and movMF, both crisp and hard, are quite large.

\begin{table}[h!]
\renewcommand{\arraystretch}{1.5}
\centering
\caption{Adjusted Rand index (ARI) and Rand index (RI) between DBMCA and SKM, between the DBMCA and the mixtures of von Mises-Fisher (crisp clastering, VMFc) and between SKM and VMFh partitioned by the number of clusters $k$. Normalized degree of concordance (NDC) and adjusted concordance index (ACI) between DBMCA and mixtures of von Mises-Fisher (soft clustering, VMFs) and between SKM and VMFs partitioned by number of clusters $k$.}
\label{tab:ARIS}
\scriptsize
\setlength\tabcolsep{2pt}
\begin{tabular}{c|cc|cc|cc|cc|cc}
\multirow{2}{*}{k} & \multicolumn{2}{c|}{DBMCA - SKM} & \multicolumn{2}{c|}{DBMCA - movMFh} & \multicolumn{2}{c|}{DBMCA - movMFs} & \multicolumn{2}{c|}{SKM - movMFh} & \multicolumn{2}{c}{SKM - movMFs} \\ \cline{2-11}
& \multicolumn{1}{c}{ARI} & \multicolumn{1}{c|}{RI} & \multicolumn{1}{c}{ARI} & \multicolumn{1}{c|}{RI} & \multicolumn{1}{c}{ACI} & \multicolumn{1}{c|}{NDC} & \multicolumn{1}{c}{ARI} & \multicolumn{1}{c|}{RI} & \multicolumn{1}{c}{ACI} & \multicolumn{1}{c}{NDC} \\ \hline
1 & 1.0000 & 1.0000 & 1.0000 & 1.0000 & 1.0000 & 1.0000 & 1.0000 & 1.0000 & 1.0000 & 1.0000 \\
2 & 0.1985 & 0.5996 & 0.1432 & 0.5725 & 0.1505 & 0.5761 & 0.6596 & 0.8321 & 0.6753 & 0.8398 \\
3 & 0.5437 & 0.7942 & 0.3735 & 0.6840 & 0.4998 & 0.7718 & 0.3826 & 0.6883 & 0.7336 & 0.8797 \\
4 & 0.5652 & 0.8308 & 0.5979 & 0.8294 & 0.5077 & 0.7850 & 0.5131 & 0.7977 & 0.4519 & 0.7647 \\
5 & 0.4625 & 0.8117 & 0.5170 & 0.8251 & 0.5865 & 0.8404 & 0.5309 & 0.8331 & 0.4344 & 0.7847 \\
6 & 0.4673 & 0.8373 & 0.5455 & 0.8355 & 0.3356 & 0.7683 & 0.4501 & 0.8073 & 0.5309 & 0.8423 \\
7 & 0.4915 & 0.8559 & 0.4537 & 0.8272 & 0.3936 & 0.8133 & 0.4802 & 0.8445 & 0.5811 & 0.8785 \\
8 & 0.4977 & 0.8710 & 0.3814 & 0.8284 & 0.4649 & 0.8482 & 0.5751 & 0.8895 & 0.5537 & 0.8809 \\
9 & 0.4645 & 0.8770 & 0.3507 & 0.8440 & 0.3386 & 0.8403 & 0.5744 & 0.8990 & 0.5676 & 0.8969 \\
10 & 0.5181 & 0.9067 & 0.3694 & 0.8637 & 0.4116 & 0.8714 & 0.5830 & 0.9088 & 0.5685 & 0.9046 \\
11 & 0.5201 & 0.9100 & 0.3419 & 0.8649 & 0.3966 & 0.8801 & 0.4954 & 0.8962 & 0.4842 & 0.8973 \\
12 & 0.5304 & 0.9166 & 0.3869 & 0.8868 & 0.3546 & 0.8787 & 0.5050 & 0.9084 & 0.5030 & 0.9064 \\
13 & 0.4365 & 0.9064 & 0.3605 & 0.8878 & 0.4030 & 0.8913 & 0.5282 & 0.9212 & 0.4263 & 0.9003 \\
14 & 0.4425 & 0.9166 & 0.3760 & 0.8960 & 0.3480 & 0.8833 & 0.5016 & 0.9202 & 0.4593 & 0.9066 \\
15 & 0.4448 & 0.9189 & 0.3858 & 0.9090 & 0.3724 & 0.9001 & 0.5849 & 0.9399 & 0.5076 & 0.9233 \\ \hline
\end{tabular}
\end{table}

%------------------------------------------------

\section{Conclusions}
\label{sec:conclusions}

We have proposed a non-parametric procedure (hence not dependent upon any distribution models) for clustering directional data. Specifically, we exploit the concept of angular data depth, which provide a measurement of the centrality of an observation within a distribution of points, to measure the similarity among spherical objects. The method is flexible and applicable even to high dimensional data sets as it is not computational intensive.

We evaluated the performances of the proposed method through an extensive simulation study. Results highlight that, when data are quite structured, it is able to recover the partitions quite well. In case of either extreme overlap (high noise) or non-structured data (that is when theoretical partitions are not governed by a clear structure), the recovery of the partitions is not good, as expected. 

In addition, we have compared our method with some other clustering techniques in the existing literature by means of a real data example about the analysis of textual data. Here again, results provide
strong empirical support for the effectiveness of our approach.

Overall, this work reveals many potential lines of work to consider in the future and more research is needed to further consolidate this interesting framework and explore its broad applications. For instance, it will be interesting to investigate its robustness properties and develop a tool to determine the number of hyper-spherical clusters using the depth method.

\bibliographystyle{unsrtnat}
\bibliography{biblio}  

\end{document}